\documentclass[conference]{IEEEtran}
\IEEEoverridecommandlockouts
\usepackage{amsmath,amssymb,amsfonts}
\usepackage{algorithmic}
\usepackage{graphicx}
\usepackage{textcomp}
\usepackage{xcolor}
\usepackage{hyperref}	
\usepackage{comment}
\usepackage{csquotes}
\usepackage{pgfgantt}
\usepackage{tikz}
\usepackage{csvsimple}
\usepackage{paralist}

\usepackage{booktabs}
\hypersetup{colorlinks, citecolor=black,
   		 	filecolor=black, linkcolor=black,
    		urlcolor=black}
\usepackage[style=ieee]{biblatex} % style=ieee or style=apa also possible

\begin{document}

%\title{Trust my fire -- Integrating software TPMs into AWS Firecracker
\title{Putting a Padlock on Lambda -- Integrating vTPMs into AWS Firecracker
%\title{Enabling Trusted PaaS -- Integrating vTPMs into AWS Firecracker
% \thanks{WASP \& Scionova AB}
}

%\author{\IEEEauthorblockN{Victor Morel}
%\IEEEauthorblockA{\textit{dept. name of %organization (of Aff.)} \\

\author{
\IEEEauthorblockN{Melker Veltman}
\IEEEauthorblockA{
% \textit{Computer Science and Engineering} \\
\textit{Chalmers University of Technology}\\
Gothenburg, Sweden \\
melkerv@chalmers.se}
\and
\IEEEauthorblockN{Alexandra Parkegren}
\IEEEauthorblockA{
% \textit{Computer Science and Engineering} \\
\textit{Chalmers University of Technology}\\
Gothenburg, Sweden \\
alepar@chalmers.se}
\and
\IEEEauthorblockN{Victor Morel}
\IEEEauthorblockA{
% \textit{Computer Science and Engineering} \\
\textit{Chalmers University of Technology}\\
Gothenburg, Sweden \\
morelv@chalmers.se}
}

\maketitle
\begin{abstract}
When software services use cloud providers to run their workloads, they place implicit trust in the cloud provider, without an explicit trust relationship.
One way to achieve such explicit trust in a computer system is to use a hardware Trusted Platform Module (TPM), a coprocessor for trusted computing. 
However, in the case of managed platform-as-a-service (PaaS) offerings, there is currently no cloud provider that exposes TPM capabilities. 
In this paper, we improve trust by integrating a virtual TPM device into the Firecracker hypervisor, originally developed by Amazon Web Services. 
In addition to this, multiple performance tests along with an attack surface analysis are performed to evaluate the impact of the changes introduced. 
We discuss the results and conclude that the slight performance decrease and attack surface increase are acceptable trade-offs in order to enable trusted computing in PaaS offerings. 
\end{abstract}
\section{Introduction}

% \todo[inline]{General introduction to trusted computing in the cloud}
Cloud computing has transformed the way we store, access, and process data.
The ability to access and use computing resources on demand, without the need for large upfront investments in hardware and infrastructure, has made cloud computing an attractive and flexible option for businesses of all sizes~\cite{surbiryala2019cloud}.
%Another purpose of using cloud technologies is the ability to hand over some management to the cloud provider, reducing the user's responsibility.
%\textcite{pichan2015cloud} argue for the relevance and importance of cloud computing, how it has changed the way Internet services are consumed, its economic opportunities, and how cloud-based solutions are expected to see greater adoption in the future.

By transferring on-site infrastructure to cloud computing platforms, more control and trust is given to the third-party company hosting the service. 
In practice, companies let cloud providers execute their software and process potentially sensitive data of their customers. 
To secure both the system running and the handling of data, safe storage and verification of the platform should be implemented to ensure trust properties.
%The potential adversary in this case is both the cloud provider itself and other customers of the cloud provider who execute code on the same physical hardware.

One way to achieve trust in a computer system is to use a \textit{Trusted Platform Module} (TPM), which is a coprocessor for secure cryptographic support. 
%The TPM interfaces with the CPU through a specific set of instructions and can authenticate the environment in which the application is executed. 
%In the case of on-site infrastructure, a TPM is implemented in hardware. 
Considering cloud infrastructure, the customer is most often given a \textit{virtual machine} (VM), which can be equipped with a \textit{virtual TPM} (vTPM). 
This vTPM is also considered trusted, as there are multiple ways to associate such a TPM with trusted hardware~\cite{vtpm}.
However, in the case where cloud providers maintain both the machine as well as the operating system, there is no one who exposes trusted computing functionality, such as a TPM.

% \todo[inline]{How does trusted computing articulates with Firecracker in particular}
Virtual machines powered by the Firecracker~\cite{firecracker-paper} \textit{Virtual Machine Monitor} (VMM), also called hypervisor in this paper, do not offer TPM device functionality.
Yet, Firecracker is used in offerings by \textit{Amazon Web Services} (AWS), the main cloud provider in the world as of today~\cite{zhang_top_2023}.
% \todo[inline]{add source here}
Firecracker was created with the goals of workload isolation, low startup times, and low memory overhead. 
To achieve minimal overhead, several trade-offs have been made compared to traditional VMMs. 
Typically, there is only support for a minimal device model, therefore not including TPM device support.
As a result, incorporating a TPM in such an environment might have an impact on said performance properties, such as memory overhead and startup times.
% To ensure that the changes introduced adhere to the goals of Firecracker, both the memory overhead and startup times are benchmarked. 

By extending the Firecracker device model, the communication between the VM and hypervisor internals is also increased. 
Such a modification can have an impact on the attack surface, which would deteriorate workload isolation.

% To assert that the changes still adhere to the goal of workload isolation for the purpose of multinenancy, we analyse the changes in the attack surface.
%The nature of the workloads also means that Firecracker has to be able to handle sudden bursts of computation, such as multiple VMs starting up at the same time. 

%Since one of the goals of Firecracker is to be able to dynamically change workloads~\cite{firecracker-paper}, software TPMs must be allocated to VMs as they are created. 
%This allocation poses a potential performance decrease in the startup time.

In order to tackle the challenges aforementioned, we address in this paper the following research questions:
\begin{enumerate}
  \item How does the startup time and memory overhead of a Firecracker VM change if a vTPM is created and allocated to it?
  \item What impact on the startup time does maintaining a resource pool of TPMs to be allocated have?
  \item How does incorporating a TPM change the attack surface of a lightweight VM?
\end{enumerate}

It is important to note that while this paper does not present a complete solution for TPMs in \textit{platform-as-a-service} (PaaS) offerings, it serves as a crucial building block towards the development of a trusted PaaS environment. 
Trusted computing already exists within \textit{infrastructure-as-a-service} (IaaS) offerings which this paper builds upon to contribute to the field through the following key aspects:
\begin{itemize}
\item We introduce TPM device support in Firecracker, ensuring the compatibility and goals of Firecracker.
\item We address the resource pooling requirements for TPMs in PaaS environments, to act as a base for future efficient and scalable implementations.
\item Our work includes a comprehensive security analysis of vTPM usage, highlighting the strengths and potential vulnerabilities associated with its implementation.
\end{itemize}
By focusing on these specific contributions, our research aims to provide valuable insights and advancements towards trusted PaaS offerings, promoting secure and reliable cloud computing platforms.

The rest of this paper is organised as follows. 
Section~\ref{related_work} reviews related concepts whilst Section~\ref{des_impl} describes how the TPM support is integrated into Firecracker.
Section~\ref{bench_resul} describes the design of the test suite and then presents the results of the performance tests and attack surface analysis.
In Section~\ref{discussion}, the results and possible impacts are discussed, and finally, Section~\ref{conclusion} concludes the paper.
\section{Related Work}\label{related_work}
This chapter is dedicated to explaining the current state-of-the-art in the space of trusted cloud computing and virtualisation technologies within Paas offerings. 
%It includes research regarding specific systems for attesting integrity in cloud environments and a discussion on recent software implementations of TPMs. 
%It also includes the integration of trusted computing primitives, such as TPMs, in cloud environments.

\subsection{VTPM Implementation Models}
To improve performance, avoid data leakage and isolation issues in vTPMs, \textcite{svtpm_a_2019} continued the work of \textcite{vtpm} and created an alternative TPM software implementation, called \textit{Secure vTPM} (SvTPM). %, not to be confused with swtpm. 
The novelty of their work is to base the trust in a TEE instead of a hardware TPM. 
More specifically, they use Intel SGX to run or protect specific parts of the TPM in an SGX enclave.  
%The result of their work is a solution where the startup time is decreased by multiple orders of magnitude in comparison to previous implementations.
Their work resulted in a much lower startup time in comparison to previous implementations.
Similarly, but more cross-platform adapted, initiatives such as HyperEnclave~\cite{jia_hyperenclave_2022} use the integrity measurement and attestation capabilities of a TPM to extend the chain of trust to the TEE.
%However, since Intel SGX is a proprietary Intel technology, it may not be applicable to cloud computing environments. 
%With this in mind, initiatives such as HyperEnclave~\cite{jia_hyperenclave_2022} can be used to make such a solution cross-platform. 
%HyperEnclave uses the integrity measurement and attestation capabilities of a TPM to extend the chain of trust to the TEE.

Another approach to achieve similar goals is called \textit{Confidential Computing TPM} (CoCoTPM).
%CoCoTPM minimises the trust required towards the host and the hypervisor, in a virtualised environment. 
%In the article, \textcite{pecholt_cocotpm_2022} apply their methods in cloud environments to support confidential computing needs. 
CoCoTPM minimises the trust required towards the host and the hypervisor by running a vTPM in an encrypted VM using AMD SEV.
The communication between these TPMs and the other VMs running the desired workload is encrypted and integrity protected, further increasing the confidentiality property%of their work
~\cite{pecholt_cocotpm_2022}.

However, to keep our contributions applicable to practical usecases, \textit{swtpm}~\cite{vtpm} is used in the experiments of this work. 
Swtpm is used by hypervisors such as QEMU and Xen.

%Contrary to the goals of a vTPM, security issues in the implementation could be an entry way for adversaries.
%A recent vulnerability was discovered in which the lack of a proper length check and proper size handling could lead to an out-of-bounds memory read and write.
%The out-of-bounds memory write vulnerability could also lead to arbitrary code execution~\cite{mitre_corporation_cve_2023,mitre_corporation_cve_2023-1,falcon_vulnerabilities_2023}.
%Taking the vulnerability into account with regards to vTPM, SvTPM and CoCoTPM highlights the need of applying defence-in-depth principles, as an adversary potentially could execute code within the environment that the vTPM is executing. 
%In the case of using a vTPM, the environment would be the host, in the case of using a SvTPM, the answer is not clear and would need to be researched further. 
%Considering the use case for a CoCoTPM, the environment would be the encrypted VM where the TPMs of the other workloads on the same host would reside, which can be considered quite critical.
%However, as these TPMs are implemented in software, the process of applying security patches is shorter compared to hardware TPMs.

\subsection{Trust in Cloud Computing Environments}
In a literature study on trust in cloud computing, \textcite{AMIbrahim2019} highlight the significant role of the TPM.
They investigate different architectures and managers for IaaS services, compare different vTPMs and remote attestation types, and discuss secure boot and integrity monitoring. 

%An interesting discussion and proposal are made by \textcite{tow_tru_2009}, where they describe a design of a \textit{trusted cloud computing platform} (TCCP). 
%The work discusses a system for trusted cloud computing through a distributed approach that matches the requirements and resources available at CSPs. 
%They use remote attestation in a distributed network and stand on previously researched primitives regarding confidential computing to hide the data of a VM to the host OS~\cite{murray_improving_2008}.

Considering PaaS offerings, \textit{Cloud Service Providers} (CSPs) are required to make a decision whether to use containerisation or virtualisation.
%, where virtualisation in this sense does not include OS-level virtualisation. 
%Both options come with shortcomings and strengths, with virtualisation being more heavyweight but well isolated, and containerisation more lightweight but less isolated. 
This comparison has seen a lot of research, both on performance and isolation~\cite{felter_updated_2015,chae_performance_2019,tesfatsion_virtualization_2018,morabito_hypervisors_2015}.

Recent development within the space of runtimes for PaaS offerings has been seen with Google gVisor from the container side, and AWS Firecracker from the virtualisation side. 
\textcite{young2019true} conducted a study comparing the gVisor runtime and the default Docker runtime.
Their results %are noteworthy, as they 
show a significant decrease in performance with respect to system calls, memory allocations, and networking. 
Another comparison between Firecracker and gVisor made by \textcite{anjali_blending_2020}, resulted in favour of Firecracker for network bandwidth, memory allocation times, and file access.

%Firecracker itself is not alone in the area of lightweight virtual machines, but takes inspiration from Solo~\cite{zhang_solo_2009} and LightVM~\cite{manco2017my}.
%Solo~\cite{zhang_solo_2009} takes a low-level approach and divides the hardware into multiple logical partitions, prioritising performance rather than isolation.
%As the VMM in Solo allows privileged instructions to execute directly on hardware, it is not applicable to multi-tenant cloud computing environments. 
%The other work describes LightVM~\cite{manco2017my}, a redesign of Xen, and achieves a major decrease in boot times, given that the workload is customised for the runtime, such as a unikernel~\cite{madhavapeddy_unikernels_2013}. 
%A unikernel is a purpose-built kernel for a specific application, in comparison to Linux, which is considered general purpose. 

%With these works in mind, the model and architecture of Firecracker is more similar to that of traditional VMMs, such as QEMU, however with a limited device model. 
%Firecracker~\cite{firecracker-paper} does not support arbitrary operating systems, but has support for both Linux and OSv~\cite{kivity_osv_2014} VMs. 
%OSv is an operating system optimised for VMs, but does run Linux applications. 

If we consider trusted computing in PaaS offerings, efforts on integrity-verified containers can be applied to solutions such as gVisor. 
An example of such work is Container-IMA by \textcite{luo2019container}.
The attestation mechanism they propose improves privacy of the container and the underlying host and ensures container isolation.
Instead of including support for vTPMs in user space, the mechanism uses a shared measurement agent in kernel space, where all application layers are measured by the IMA.
The novelty of their work lies in their method for multiplexing PCRs for ephemeral container workloads.

However, for PaaS solutions based on virtualisation, such as Firecracker, no effort has been made to enable trusted or confidential computing.
As our work aims to integrate TPMs into Firecracker, other components are needed to provide trust, confidentiality, and integrity to PaaS offerings.
Already available components, such as the mentioned SvTPM~\cite{svtpm_a_2019} and Linux IMA~\cite{sailer_ima_2004} can be used alongside our contributions to build a complete system. 

\section{Design and Implementation}\label{des_impl}
In this section, we introduce the changes made to Firecracker to support a vTPM.

\subsection{High-level Description}
%The modified Firecracker VMM needs the ability to pass TPM devices to guests. 
The current virtual device capabilities of Firecracker are small, focusing on devices adhering to the virtio standard~\cite{russell_virtio_2008}.
The TPM is usually discovered using \textit{Advanced Configuration and Power Interface} (ACPI), but to fit the current Firecracker device model that does not support ACPI, the virtual device implemented in this project uses virtio.
After analysing different VMMs, only crosvm~\cite{crosvm_2023} uses TPM over virtio, in comparison to QEMU and Cloud-Hypervisor, which both use ACPI~\cite{bellard_qemu_2005,cloud-hypervisor}.

\begin{figure}[h]
    \centering
    \includegraphics[width=80mm]{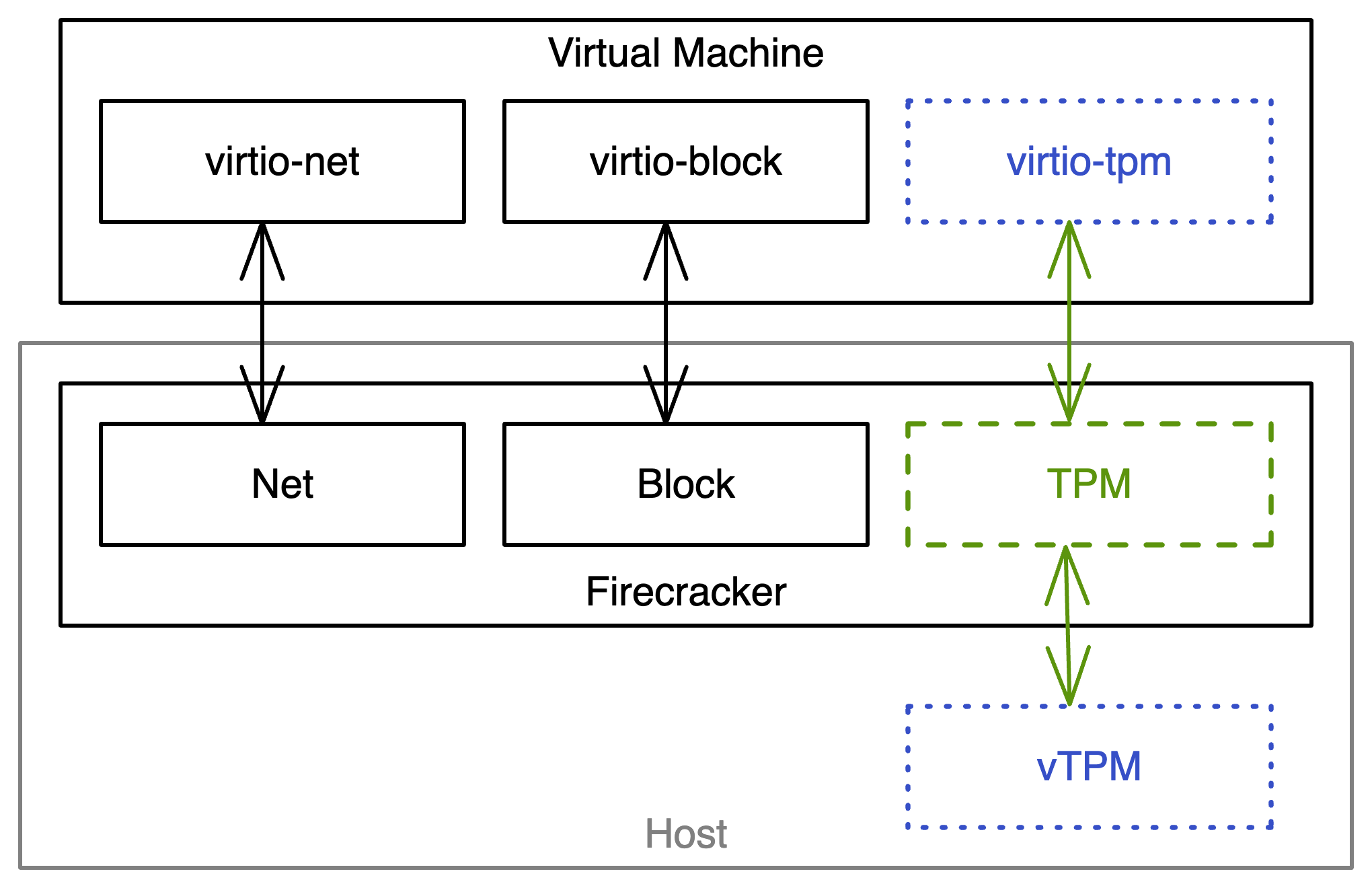}
    \caption[Firecracker Device Model with TPM Support]{The device model of Firecracker with TPM device support}
    \label{fig:tpm-additions}
\end{figure}

The updated Firecracker device model can be seen in \autoref{fig:tpm-additions}. 
The boxes with blue, dotted outlines denote the added components from open-source projects, whereas the boxes with green, dashed outlines are the components implemented in this project. 
The boxes inside the Firecracker box represent device interfaces between the host and the guest.
%The vTPM started on the host OS is connected to the Firecracker VMM through a UNIX socket. 
The specific vTPM implementation on the host used in this project is swtpm~\cite{vtpm}.
%A UNIX domain socket provides interprocess communication with semantics similar to a network socket.
%The implementation of the virtual TPM device in Firecracker exposes the device according to the virtio standard~\cite{russell_virtio_2008} to be consumed by the guest. 
The virtio-tpm device driver on the guest side uses the already existing architecture for TPMs in the Linux kernel and integrates with the IMA~\cite{tolnay_chromium_nodate}.

%On the guest side, there is currently no virtio TPM device driver in standard Linux. 
%However, in a Linux fork by Google, such a driver is present. 
%As that driver is open source, it can be used for this purpose~\cite{tolnay_chromium_nodate}.
%This device driver uses the already existing architecture for TPMs in the Linux kernel and integrates with the IMA.

To run a Firecracker VM with a TPM, some cryptographic certificates and keys need to be created to associate the vTPM with the hardware TPM on the host. 
After that, the vTPM should be started with a file path passed to it. 
The vTPM will then create a communication socket on that file path. 
This socket then needs to be passed to the Firecracker VM configuration, along with a Linux kernel compiled with virtio-tpm support.
Then the Firecracker VM can be started and will have access to the vTPM.

\subsection{Technical Implementation}
\begin{figure}[h]
    \centering
    \includegraphics[width=80mm]{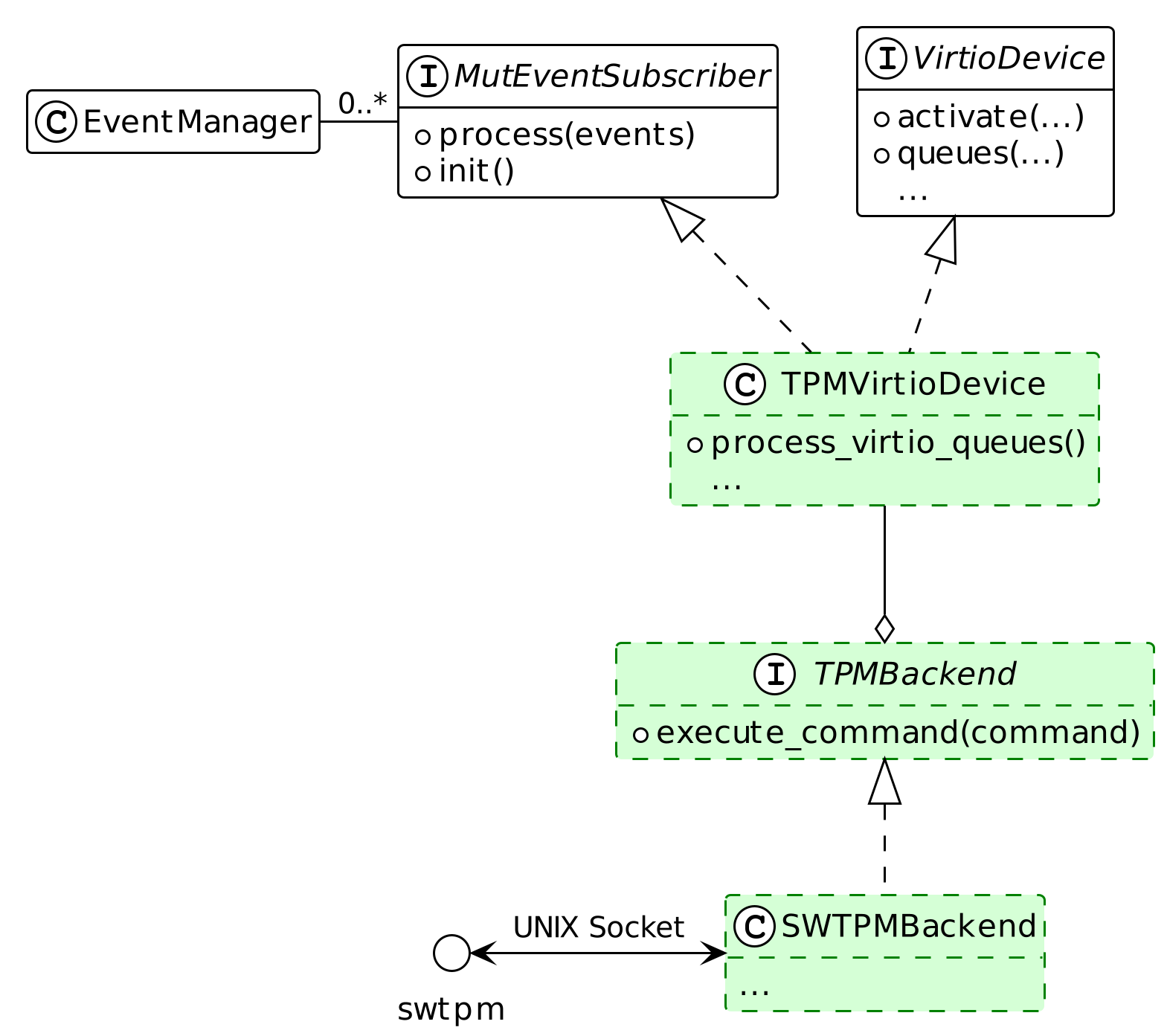}
    \caption[Class Diagram of Firecracker Changes]{An UML class diagram, the green boxes with dashed borders are added implementations to Firecracker}
    \label{fig:firecracker-tpm-class-changes}
\end{figure}

The implementation of the TPM device in Firecracker can be seen in \autoref{fig:firecracker-tpm-class-changes}. 
The two main components can also be seen in \autoref{fig:tpm-additions} as the two green arrows.
\emph{SWTPMBackend} is the arrow from the vTPM to the TPM and \emph{TPMVirtioDevice} can be seen as the arrow between the TPM box and the virtio-tpm box.

\begin{comment}
@startuml Firecracker TPM Implementation
!theme plain

class TPMVirtioDevice #d5ffd6 ##[dashed]green implements VirtioDevice, MutEventSubscriber  {
    + process_virtio_queues()
    {method} ...
}
interface VirtioDevice {
    + activate(...)
    + queues(...)
    {method} ...
}
interface MutEventSubscriber {
    + process(events)
    + init()
}
interface TPMBackend #d5ffd6 ##[dashed]green {
    + execute_command(command)
}

class EventManager

class SWTPMBackend #d5ffd6 ##[dashed]green implements TPMBackend {
    {method} ...
}

() swtpm
swtpm <-> SWTPMBackend : UNIX Socket
TPMBackend o-- TPMVirtioDevice
EventManager "0..*" - MutEventSubscriber

hide empty members
@enduml
\end{comment}

The \emph{SWTPMBackend} was inspired by the vTPM implementation in the Cloud-Hypervisor project~\cite{cloud_hypervisor_2023}, as they also use swtpm over a UNIX socket.
To account for platform setup procedures, swtpm exposes a control channel, which is used to both initialise the TPM and to set a file descriptor for client TPM commands.
The file descriptor is one side of another UNIX socket that is created in \emph{SWTPMBackend}. 
The other side of that socket is used to send and receive messages inside Firecracker. 

The \emph{TPMVirtioDevice} implements the \emph{MutEventSubscriber} and \emph{VirtioDevice} interfaces which already exists in Firecracker.
By using these interfaces, the virtual device fits into the Firecracker device model, and no major changes were needed regarding the internals of Firecracker. 
Another major advantage of implementing these interfaces was that interrupt registration and device activation are taken care of by already existing components within Firecracker.
Firecracker also passes the memory address of the device automatically to the kernel command line parameters for device discovery by the Linux kernel. 

In the virtual device implementation, a single queue is used for communication between the VM and Firecracker. 
Using a single queue was possible as it is always the client who initiates communication in the case of a TPM, which can be compared with a network device where any side might initiate communication. 
This difference is also apparent when comparing the virtio network device in Firecracker with our virtio TPM device, as the virtio network device uses two queues. 

The source code for the virtual TPM device, as well as the resources required to reproduce the benchmarks described in the next section, can be found in a GitHub repository~\footnote{\url{https://github.com/DATX05-CSN-8/fctpm-2023/tree/paper}}.
\section{Benchmark and Results}\label{bench_resul}
In order to answer our research questions, we describe how our performance tests operate and present the results of these tests in this section.
We also analyze the changes in the attack surface.

\subsection{Design of the Test Suite}
%\section{Test Environment Setup}
\label{perftest-hardware}
The performance tests are performed using CloudLab, which is a scientific infrastructure that gives bare metal access to cloud computing resources~\cite{duplyakin_design_cloudlab_2019}.  
More specifically, the tests run on an instance consisting of two Intel Xeon Gold 6142 with 16 cores each and 384 GB of memory.
%To minimise the setup for the performance test instances, pre-compiled binaries from another machine are used instead of compiling the binaries on the performance test instances. Along with this, software packages for swtpm were installed. 

%The hardware environment consists of two different setups. 
%One for the development and functional tests of the project and one for executing the performance test suite. 
%The hardware environment during the development phase was an x86-64 computer with a physical TPM, capable of virtualisation. 
%In the performance test phase, the system defined in \autoref{perftest-hardware} was used.

%To verify the hardware environments, two tests were performed. 
%The first test had the purpose of verifying the virtualisation capabilities of the environment.
%To perform this test, a pre-built Firecracker binary was used along with a Linux distribution created by LinuxKit, a toolkit for building reproducible Linux distributions. 
%The assertion was that the kernel started up correctly. 
%The second test was intended to verify the vTPM functionality in a VM. 
%This test was done using QEMU, swtpm and a Linux distribution created by LinuxKit. 
%The assertion was that the TPM could be interfaced with from the virtual machine. 

%In addition to the tests, software packages were installed to enable the building of Firecracker and the necessary components. 

%\section{Performance Test Suite}
\label{method-perf-test-suite}
The performance test suite runs multiple scenarios with multiple setups. 
The first setup, \emph{baseline} runs an unmodified Firecracker binary without a TPM, to serve as a comparison value. 
The second setup, \emph{on demand} runs the modified Firecracker binary and provisions keys and starts a vTPM right before each VM is started.
The third setup, \emph{pool} maintains a resource pool of already provisioned vTPMs and allocates one of them to each Firecracker VM before starting the VM. 

Regarding the VM OS used, a Linux kernel compiled with a virtio-tpm driver along with a simple init script is used.
The init script also shuts down the VM directly.
The reason to not start any actual service is to minimise the number of components that can impact the startup time.
The same OS configuration is used for all startup time tests. 

The scenarios executed are the following:
\begin{itemize}
    \item 500 VMs, serially executed
    \item 1000 VMs, 50 running concurrently
\end{itemize}
These scenarios are the same as those used in benchmarks made comparing Firecracker with QEMU and Cloud-Hypervisor~\cite{firecracker-paper}.

In addition to the startup time metric, we also measured the memory overhead of adding the TPM to the VM.
%As the VM needs to be kept running to be able to analyse the memory, it uses a different init program, which spawns a shell. 
%When the memory measurement is finished, the VM will be manually terminated from the host.
The memory overhead describes the memory usage of each VM, disregarding the memory actually available inside the VM. However, different VM memory sizes were used to verify that the memory overhead was stable.
Memory overhead is measured using the \emph{pmap}~\cite{cahalan_pmap1_nodate} tool, which displays Linux process memory maps.
Only non-shared memory is counted, as it represents how the system scales with more VMs.
Linux allocates the same physical memory region for processes using the same read-only memory, therefore, it does not have an impact on a system with thousands of VMs. 
The memory overhead benchmark evaluates 3 setups: baseline, modified excluding vTPM memory, and modified including vTPM memory.
%Three different setups are used here as well; it includes a VM with an unmodified Firecracker binary, as well as two setups with modified Firecracker binaries. 
%The two modified setups differ in whether or not the memory of the swtpm processes is counted. 

The memory of the TPM pool is not considered, as it is not part of either the Firecracker process or the vTPM process. 
%The purpose of the TPM pool implementation in this project is to isolate the boot time from the vTPM startup time.
In a real-world scenario, the pool itself would be implemented by the cloud provider to optimise for their use case. 
%Due to this implementation, the pool does not need to be considered for the memory overhead results.

\subsection{Pooling Algorithm}
\label{pooling_algorithm}
For the case of running VMs with TPMs allocated from a pool, the pool adheres to a specific resource pooling algorithm. 
The purpose of using a pool is to move the resource allocation time from a critical point in time to a time where resources and time are less scarce. 
A characteristic of a vTPM resource is that it can be used by only one VM throughout its lifecycle, which becomes a strict requirement of the resource pool. 
The algorithm implemented for this project is a simple pre-allocated buffer, where vTPMs are provisioned and started when the pool is created.

\subsection{Results}
This chapter presents the test results, mainly the effects of integrating TPMs into Firecracker with regard to performance and security. 
Based on these scenarios, the results are visualised using a graphed \textit{cumulative distribution function} (CDF) of the startup times along with a table that shows a numerical interpretation of the samples.
%A CDF will be used instead of a scatter plot to better visualise the distribution of our results. 
%The data obtained by running the test suite have been parsed, and are presented in diagrams for an easier understanding and overview. 
%Regarding the security aspect, the changes to the attack surface are presented at a technical level. 
%\subsubsection{Performance overhead}
%The performance impact of integrating TPMs into Firecracker has been evaluated by measuring the changes in VM startup time and memory overhead. 
%Changes in VM startup time were evaluated with two different scenarios, 500 VMs running without concurrency and 1000 VMs with 50 VMs running concurrently using the performance test suite explained in \autoref{implementation}.
%Measuring the VM startup time was done with 3 different setups: baseline, on demand, and pool.
%The \emph{baseline} setup uses an unmodified Firecracker binary, whereas the \emph{on demand} setup and the \emph{pool} setup uses a Firecracker binary with added TPM support.
%The difference between \emph{on demand} and \emph{pool} is that \emph{on demand} allocates TPMs as they are needed, whereas \emph{pool} allocates TPMs before the VMs start. Therefore, the \emph{pool} setup removes the TPM allocation time from the startup time. 
To ensure that no measurement errors occurred, the startup time measured from the performance tests was compared with the time reported from the internal boot timer device in the Firecracker hypervisor. 
The difference between the two was of reasonable size, ranging from $10-20$\%, implying no noteworthy errors.%the trustworthiness of the results. 

\subsubsection*{Startup time scenario 1, 500 VMs, no concurrency}
\begin{figure}[h]
    \centering
    \includegraphics[width=80mm]{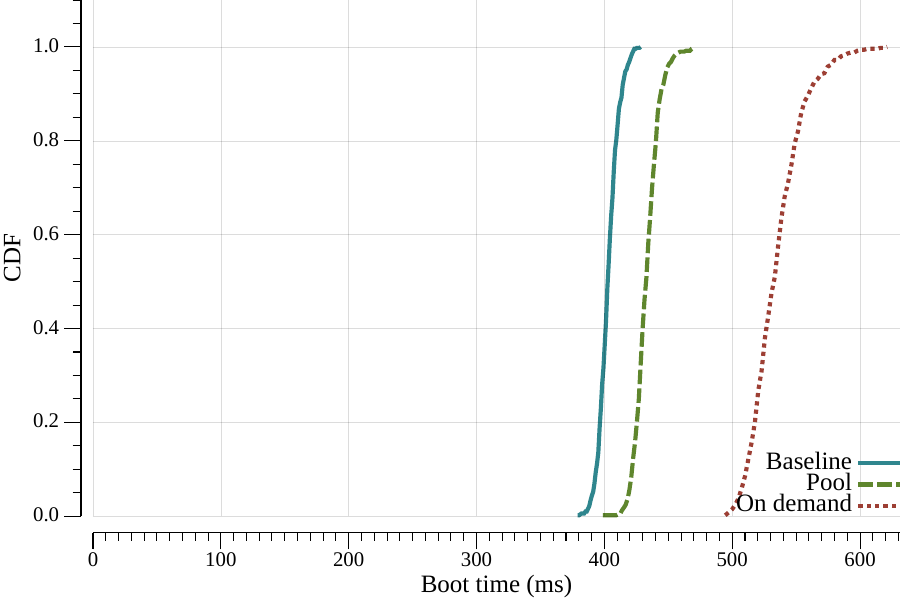}
    \caption[Distribution of Boot Times for 500 VMs]{Boot time for 500 VMs serially executed}
    \label{fig:boottime_chart_500}
\end{figure}

%\textbf{Startup time scenario 1: 500 VMs, no concurrency.} 
The results of the scenario when running 500 VMs can be seen as a chart of the CDF in \autoref{fig:boottime_chart_500} and as numerical data in \autoref{tab:boottime_table_500}. 

\begin{table}[h]
    \centering
\csvreader[
  tabular=lrrrrrr,
  table head=\toprule {Type} & {p95} & {p95 Change} & {Std. Deviation} \\\midrule,
  late after last line=\\\bottomrule\newline
]{
  figure/boottime_table_500.csv
}{}{\csvcoli & \csvcolv & \csvcolvi & \csvcolvii}
\caption[Boot Times for 500 VMs]{Boot time for 500 VMs serially executed}
\label{tab:boottime_table_500}
\end{table}

An important aspect is the standard deviation that demonstrates the width of each CDF.
As can be seen in \autoref{tab:boottime_table_500}, the difference between the baseline and the pool implementation is significantly smaller in comparison to the difference between the baseline and the on demand implementation. 
With regard to absolute numbers, the p95 column, showing the 95th percentile, tells us that the same pattern is seen here. 
%Another interesting metric is the \textit{p95} column of the table, which displays the number of samples that are below 95\% of all samples. The p95 value can be used to mitigate the impact of outliers while taking the deviation of samples into account. 
%This metric can be seen as a combination of the absolute value and the deviation, which is also seen in the value of the metric.
In general, the pool implementation performs slightly worse than the baseline; however, if the TPMs are allocated on demand, the results display a significantly worse boot time.

\subsubsection*{Startup time scenario 2, 1000 VMs, 50 concurrent VMs}
\begin{figure}[h]
    \centering
    \includegraphics[width=80mm]{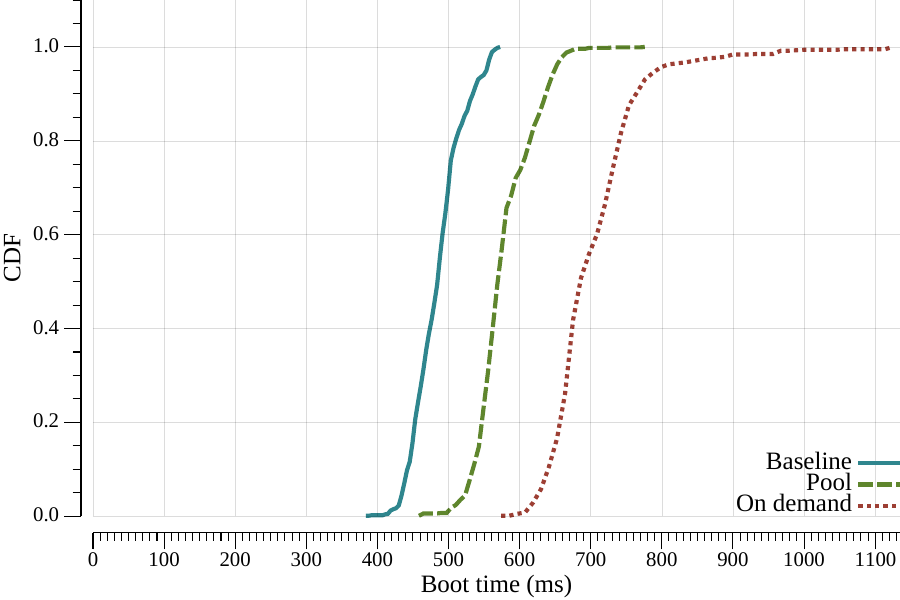}
    \caption[Distribution of Boot Times for 1000 VMs]{Boot time for 1000 VMs, with 50 running concurrently}
    \label{fig:boottime_chart_1000}
\end{figure}

Considering the other scenario, with 1000 VMs in total and 50 of them running concurrently, as can be seen in \autoref{fig:boottime_chart_1000} and \autoref{tab:boottime_table_1000}, the result changes. 
The similarity regarding the CDF of the baseline and the pool setup is less apparent, and the CDF for the pool setup is slightly wider. 
However, the pool setup still performs significantly better in comparison to the on demand setup.
An overall change is that the deviations of all three setups increase by at least three times from scenario 1. 

\begin{table}[h]
    \centering
\csvreader[
  tabular=lrrrrrr,
  table head=\toprule {Type} & {p95} & {p95 Change} & {Std. Deviation} \\\midrule,
  late after last line=\\\bottomrule\newline % horizontal line at the end of the table
]{
  figure/boottime_table_1000.csv
}{}{\csvcoli & \csvcolv & \csvcolvi & \csvcolvii}
\caption[Boot Times for 1000 VMs]{Boot time for 1000 VMs, with 50 VMs running concurrently}
    \label{tab:boottime_table_1000}
\end{table}

\subsubsection*{Memory Overhead}
%The memory overhead test was performed with multiple memory sizes allocated to the VM, to validate that there was no change in overhead dependent on the memory size allocated to the VM. 
%The results did not show any changes in overhead across VM memory sizes, which can be seen in  \appendixautorefname~\ref{appendix:memoverhead_chart}. 
%A condensed version, leaving out different VM sizes, can be seen in \autoref{tab:memoverhead_table}.
%The memory overhead measurements were also executed for different VM memory sizes, to verify that the overhead is stable. 
%We did not see any change in memory overhead for VM sizes between 128MB and 8GB.
The results showed no divergence when different VM memory sizes were used, verifying a stable overhead. 
%The results of the memory overhead measurements can be seen in \autoref{tab:memoverhead_table}.
%The two setups where only the Firecracker process was measured are almost the same.
%However, when the vTPM process was included, the impact on performance is significantly greater. 
%The increase in memory usage considering only the modified Firecracker process is $4$KB, corresponding to an $0.16$\% increase.
%However, when the memory of the swtpm process is taken into account, the change increases to $52.35$\%.
The change in memory overhead for the 3 setups can be seen in \autoref{tab:memoverhead_table}.
When only the Firecracker process was measured, the difference was only $4$KB, corresponding to a $0.16$\% increase.
However, when including the memory of the swtpm process, the impact was significantly greater showing an increase of a total of $52.35$\%.
The significance of the $3760$KB overhead is further discussed in Section~\ref{disc_memoverhead}.

\begin{table}[h]
    \centering
\csvreader[
  tabular=lrr,
  table head=\toprule  & KB Overhead & Change \\\midrule,
  late after last line=\\\bottomrule\newline % horizontal line at the end of the table
]{
  figure/memoverhead_table.csv
}{}{\csvlinetotablerow}
\caption[Memory Overhead Results]{Memory overhead of Firecracker and corresponding percentage increase}
    \label{tab:memoverhead_table}
\end{table}

\subsection{Attack Surface Analysis}
\begin{comment}
\begin{table}[h]
    \centering
\csvreader[
  tabular=lrrrrr,
  table head=\toprule Language & Unchanged & Changed & Added & Removed & Difference \\\midrule,
  late after last line=\\\bottomrule % horizontal line at the end of the table
]{
  figure/loc_table.csv
}{}{\csvlinetotablerow}
\caption[Changes in Lines of Code]{Lines of code edited by adding TPM support to Firecracker}
    \label{tab:loc_diff}
\end{table}
\todo[inline]{probably remove lines of code change}

As Firecracker aim to have a minimal implementation, there is relevancy in understanding what impact the modifications have to understand the difference in regards to the attack surface. 
The difference from the modified Firecracker VMM compared to the original code, in terms of the number of lines of code, has been examined using the command line program CLOC~\cite{aldanial_cloc_2022}. 
The scope of the modification made to the Firecracker VMM can be seen in \autoref{tab:loc_diff}.
Considering only Rust files as program code, since YAML, TOML, and JSON is configuration, the changes resulted in an increase of $2$\% lines of code. 
\end{comment}
As Firecracker aim to have a minimal implementation, there is relevancy in understanding what impact the modifications have.
The modified Firecracker VMM compared to the original code in terms of the number of lines of code, was examined using the command line program CLOC~\cite{aldanial_cloc_2022}. 
Considering only program code, disregarding configuration files, there was only an increase of approximately $2$\% lines of code (which corresponds to an addition of 1257 lines for a total of 61 697).
% 1 257 lines added compared to total 61 697

% added 1 257
% removed 12
% unchanged  61695
% changed 2

%As Firecracker already has general virtual device support, there was no change in attack surface with regard to the communication between a guest and the hypervisor code. 
%What is needed to be analysed further is how the hypervisor internals communicates with other components in the host OS. 
%A reasonable comparison can be made between the virtual network device and the virtual TPM device in Firecracker. 
%The virtual network device creates a simulated network device using Linux kernel functionality. 
%It is created using a function call to a Linux kernel module and therefore executes kernel code based on the configuration supplied to Firecracker.
%As the network device is implemented in a kernel module, all network calls from within the guest will also pass through the kernel. 
%This procedure can be compared with the virtual TPM device that communicates with swtpm, another userspace process on the host, therefore being significantly less privileged. 
%However, the virtual TPM device uses a UNIX socket to communicate with the swtpm process, and the UNIX socket implementation reside in the kernel.
%However, this connection is not central for the TPM functionality, but is used as a communication medium. 
The communication between a guest and the hypervisor code had no change in the attack surface, considering Firecracker already handles virtio support.
Instead, we made a comparison between the network device and the TPM device, regarding their communication between the hypervisor internals and the host OS.
The virtual network device creates a simulated network device using a call to a Linux kernel module that executes code based on the configuration supplied to Firecracker.
Therefore, all network calls in the guest will pass through the kernel. 
This procedure can be compared with the TPM device in Firecracker that communicates with swtpm -- a userspace process on the host --, therefore being significantly less privileged. 
%However, the communication uses a UNIX socket which is implemented in the kernel. 
%But this connection is not central for the TPM functionality and is only used as a communication medium. 

Seeing that the vTPM device in Firecracker passes the data from the guest to the swtpm process itself, it should be considered an extension of the attack surface. 
Swtpm, implemented in C, poses a potential target for adversaries due to memory safety issues, as can be seen in recent issues~\cite{falcon_vulnerabilities_2023}.
Such issues can lead to code execution vulnerabilities, which in the case of an swtpm process interfaced from a Firecracker VM, implies a VM escape. 
%A VM escape is when a user within the VM has achieved code execution abilities on the host. 
However, since the swtpm process runs in the host userspace, defence-in-depth principles can be used to isolate the process from the rest of the system. 

These findings clearly indicated that the greatest change in the attack surface originates from the swtpm program itself and not from the integrated TPM functionality in Firecracker.

\section{Discussion}\label{discussion}
This section discusses the results of the experiments, measurements, and analyses performed, with respect to both performance and security. 
The impact on performance and security introduced by adding TPM support to Firecracker is evaluated in a real-world scenario with CSPs in mind. 
Additionally, limitations and potential measurement errors are discussed. 

%The chapter ends by describing how recent research within trusted computing can be applied to further improve workload isolation with TPM support in mind. 

\subsection{Performance Evaluation}
In this section we first discuss various aspects regarding the startup time and then the results of the memory overheads.

\subsubsection{Startup Time Evaluation}
%The results for the test scenario without parallelism, considering only the pooled implementation, show an increase in startup time of $8$\%, which should be considered acceptable by both CSPs and their customer. 
%However, considering the scenario of running a total of 1000 VMs with 50 VMs running concurrently, the p95 value of the pooled setup resulted in a $17$\% increase, which is significant. 
The results for the concurrent test scenario using the pooled setup showed a $17$\% increase which might seem quite major. 
However, since the usage of a TPM in PaaS solutions would be an opt-in feature, similar to that of the current IaaS, %t would not be considered an issue for the customer, as they can decide whether to use this functionality or not.
customers will have the flexibility to choose whether or not to utilize this functionality.
%the choice to employ and utilize this functionality lies with the customers.
Additionally, the CSP could charge an additional fee for this functionality, depending on whether the performance impact hits the CSP or the customer. 

Furthermore, since the output of the Firecracker internal boot timer device was compared with the values measured by the test suite, measurement errors can be ignored. 
The startup time saw quite a major increase regarding the pooled setup moving Scenario 1 to Scenario 2. 
As measurement errors were ruled out by comparing with the internal boot timer, we can consider the communication introduced between the Firecracker hypervisor and the vTPM as a root cause. 

%\subsection{Caveats in a Concurrent Environment}
The virtual TPM device implemented in Firecracker is synchronous, meaning that the hypervisor wakes up as the guest notifies the device, after that the request is sent to the vTPM, and the guest stays paused for the time it takes to process the request.
Also, when sending and receiving data from the SWTPMBackend, the Firecracker process is blocked due to IO calls to the file descriptor communicating with the swtpm process. 
The IO calls cause the process to go to sleep; meanwhile, another process might start to execute, which can cause a delay. 
As the number of concurrent processes increases, the risk that the Firecracker process needs to wait before executing again after a blocking call increases. 

%\subsection{Differences to Previous Work}
When comparing the startup time results in this project with the values in the Firecracker research paper~\cite{firecracker-paper}, the difference is significant. 
%One potential reason behind this disparity is that the Linux kernel configuration used in this project is different. 
%The reason to use a different kernel configuration is the requirements of the virtio TPM device driver. 
%As the device driver makes use of the complete TPM device subsystem, more kernel components need to be loaded, causing an increase in startup time. 
%An option would be to use a kernel with the same configuration as used in the research paper for the baseline, which would put emphasis on the changes introduced to the Linux kernel and not to the Firecracker VMM itself. 
A reason behind the difference in startup time is the different hardware setups used, mainly the CPU used could have impacted the result. 
The CPU used by \textcite{firecracker-paper} is more performant in comparison to the one used in this project. 
Both single-thread performance and multi-thread performance are significantly lower in this project compared to the one used in the Firecracker research paper. 
Important for the concurrent scenario is the number of cores, which in the Firecracker paper totalled 48, compared to the 32 used in this project. 
An important aspect regarding the number of cores is the difference between the total number of cores and the desired parallelism. 
%Logically, if the number of cores is greater than the desired parallelism, all workloads can be executed completely in parallel. 
Logically, as the number of cores is lower in comparison with the desired parallelism, the host OS needs to make vital scheduling decisions.
These decisions may have caused some workloads to wait, which might have been a reason for the performance difference.

%\subsection{Resource Pool}
Regarding the usage of a resource pool, considering real-world use, there would not be a situation where a CSP would allocate vTPMs on demand. 
Cloud providers go to great lengths to minimise the critical execution path to improve the response times of their PaaS offerings.
Therefore, if there is an option to use a resource pool, CSPs have no reason not to use it.
However, the pool algorithm featured in this work is not suitable for real-world PaaS environments, as it assumes that the total number of virtual machines is known in advance. 
%A viable, although simple, solution would be to always instantiate a new resource and put it in the pool every time an allocation occurs. 
%More advanced solutions could make use of other metrics such as the average time between VM starts and instantiate resources accordingly in a separate thread. 

%Additionally, the memory usage of a dynamic resource pool is likely to be correlated with the number of resources needed per time unit. 
%The correlation is due to the need to hold a set of not yet allocated resources, a set that grows with the number of resources needed per time unit, which is largely dependent on the deployment method used at a CSP. 

\subsubsection{Memory Overhead Evaluation}
\label{disc_memoverhead}
With regards to the memory overhead comparing the unmodified Firecracker and Firecracker with TPM support, a static increase of 4KB was observed regardless of the VM memory size. 
Such an increase is minor and would be considered tolerable by a CSP. 
However, including the overhead of the swtpm process, the total overhead amounts to 3760KB.
%Such an increase might seem major; however, 
Which, considering the total memory allocation of a VM still corresponds to a minor part of the total memory allocated, typically a minimum of 128MB. 

\begin{comment}
\subsection{Relation to VM Memory Sizes}
Since the memory overhead is static across VM memory sizes, the ratio between the overhead and the desired VM memory decreases as the desired VM memory increases.
The ratios for the three smallest VM sizes were $2.89$\%, $1.43$\%, $0.72$\%, with the memory sizes 128MB, 256MB and 512MB, respectively.
Note that these ratios are for the modified Firecracker VMM, including the swtpm process. 
The corresponding ratios for the unmodified Firecracker VMM is $1.88$\%, $0.94$\%, and $0.47$\%. 
These numbers show that, although there is a significant increase in memory overhead, the overhead compared to the VM memory size is still minimal.
\end{comment}

%\subsection{Compared to Other VMMs}
Another aspect is to compare Firecracker to QEMU and Cloud-Hypervisor.
\textcite{firecracker-paper} stated that Firecracker have the smallest overhead at approximately 3MB, whereas Cloud-Hypervisor came second at approximately 13MB, and QEMU last at 131MB~\cite{firecracker-paper}. 
%Another comparison to take into account is the memory overhead results of the Firecracker paper. 
%In their measurements, comparing QEMU, Cloud-Hypervisor and Firecracker, Firecracker had the smallest overhead at approximately 3MB, whereas Cloud-Hypervisor came second at approximately 13MB, and QEMU last at 131MB~\cite{firecracker-paper}. 
Considering the overhead of the other hypervisors, the increase seen when adding TPM support is negligible. 

As stated, the memory overhead of the resource pool was not considered during the tests.
This is mainly due to the pooling algorithm used as it pre-allocates TPMs for the total number of VMs, which means it consumes more memory compared to a dynamically allocating algorithm. 

\subsection{Security Analysis}
%\section{Security Evaluation}
\label{disc_security_evaluation}
The result of the attack surface analysis highlights that the major change in the attack surface is seen in the swtpm process.
As it is not an option to limit the communication with the swtpm process from the Firecracker hypervisor --considering it needs to adhere to the TPM specification --, other security hardening options have to be used.
Assuming the worst, that an adversary can achieve code execution through the TPM commands to the swtpm process, applying defence-in-depth principles come naturally to isolate the adversary and minimise the impact that can be made. 
Defence-in-depth can be applied and used in multiple ways; one alternative is to isolate the process using Linux kernel primitives, another is to run the vTPM in a separate VM as done with CoCoTPM~\cite{pecholt_cocotpm_2022}.

\subsubsection{Isolating with Linux Kernel Functionality}
To harden the security of the swtpm process, Linux cgroups, namespaces, and seccomp filters can be used, among others. 
Using cgroups can mitigate host resource exhaustion if the adversary floods the swtpm process through the TPM interface. 
Namespaces improve process isolation by exposing an abstracted view of the network, the process table, and the directory structure to the environment of the swtpm process. 
Therefore, the use of namespaces can mitigate potential pivots of an adversary, as it can limit the network capability of the process.
By using seccomp filters, the \textit{system calls} (syscalls) a process can perform are limited to a specific set. 
%Syscalls are a way for a userspace process to communicate with the Linux kernel.
%Using seccomp filters therefore severely limits the ability of a process, as communication primitives within a Linux environment commonly require a syscall. 
As the vTPM emulates a self-contained hardware component with limited connectivity, few syscalls need to be allowed by the seccomp filters, showcasing the applicability for this use case.

\subsubsection{Alternative Isolation Methods}
Another option to improve the security of the vTPM implementation is to contain the emulation within the Firecracker program itself. 
Containing the emulation within Firecracker would, however, increase the program size and memory use of the Firecracker process along with coupling the two components together, which might be undesirable for a CSP. 
Also, if the swtpm code were to be part of the Firecracker process, an arbitrary code execution exploit would achieve the privilege of the Firecracker process.
Therefore, an adversary would have more access compared to the case where swtpm is a separate, hardened process.

Containing the vTPM in a separate environment is another option, which would improve isolation even more, at the price of increased overhead. 
Previous work within the space shows both solutions in separate VMs as well as TEEs~\cite{pecholt_cocotpm_2022,svtpm_a_2019}.

Although the security impact can be interpreted as major due to the risk of vulnerabilities in the vTPM implementation, the time to patch a software component is significantly shorter compared to patching a hardware implementation or firmware. 
Combining the shortened patch time with the fact that CSPs already use vTPMs in their IaaS offerings implies that the security impact is considered acceptable.

\section{Future Work}
As this work is aimed at a specific building block that enables trusted computing for PaaS offerings, there are several improvement areas and subfields that would benefit from future research. 

\subsection{VM Exits}
One metric that is neither evaluated nor analysed is the number of \emph{VM exits} occurring.
A VM exit is when control is transferred from execution within the VM to the hypervisor code, commonly occurring when a VM interacts with an emulated device. 
As VM exits can be considered an attack vector~\cite{szefer_eliminating_2011}, it would therefore be beneficial to compare the number of VM exits with and without a TPM.
VM exits require significant insight into KVM components to be able to analyse the different reasons behind the exits. 
Hence, the topic of VM exits have not been researched in this work as it deviates from the main purpose of this work.

\subsection{vTPM Trust Establishment at Scale}
Within practical applications, there is also multiple research subfields to be further worked upon.
One such area is trust processes and trust relationships, related to simulated TPM manufacturing and remote attestation. 
The concepts described in this project need to be adapted to a larger scale to be applicable for CSPs in practice. 

\subsection{Trusted Late Launch}
Efforts can also be made to implement trusted late launch functionality, where a VM can boot before it is known what workload it will execute, to later receive the workload and perform integrity measurements. 
By using a late launch, the startup time for workloads would be able to be reduced even further. 

Further testing could also be beneficial to optimise the process and overhead from starting and using a vTPM in Firecracker.
Similar to how \textcite{manco2017my} examine the overhead of short-lived lightweight VMs whilst keeping isolation properties, it would be interesting to see what impact the vTPM support can have on more applied use cases.
However, optimising the startup time would require knowledge of the vendor-specific resources and the Firecracker VMM.

\subsection{Alternative vTPM Implementation}
The vTPM implementation used is important with regard to the security of a trusted PaaS offering by a CSP. 
As mentioned in \autoref{disc_security_evaluation}, multiple actions can be taken to further isolate a Linux process to mitigate potential vulnerabilities. 
One potential research topic is to implement a vTPM in a memory-safe language, effectively protecting it against memory corruption attacks similar to those seen previously with vTPMs~\cite{mitre_corporation_cve_2023}.

\subsection{Migrations of VMs}
Another relevant research avenue is snapshots and migrations of VMs. 
Although previous work has been conducted with other VMMs regarding this topic~\cite{vtpm-vm-migration}, it falls outside the scope of this thesis and can be considered as potential future work.

\section{Conclusion}\label{conclusion}
This project has integrated software TPM support into the Firecracker VMM~\footnote{The maintainers of Firecracker have been contacted through GitHub Issues, see \url{https://github.com/firecracker-microvm/firecracker/issues/3629}.} along with measuring the performance and security impact of the additions.
By incorporating the TPM support, the memory overhead of the Firecracker process saw a negligible increase.
When including the vTPM process in the memory overhead, the memory overhead increased significantly. 
Nevertheless, the overhead remains a small percentage of the total memory allocated to a VM, and would, if used in practice, remain an opt-in feature for customers. 

The startup time increased significantly when TPMs were allocated on demand; however, when using a resource pool with pre-allocated vTPMs, the startup times were kept at an acceptable level.
Yet, there is still potential to further improve these metrics even further by adjusting the resource pool algorithm based on the workload and the specific hardware used.

Regarding the attack surface, there is an increase originating from the extended communication between the VM and external processes. 
However, considering that other virtio devices are already supported, the extended support for a virtio TPM fits into the Firecracker device model and does not extend the attack surface there. 
Therefore, the additional attack vector lies in the actual TPM implementation, which could easily be interchanged and hardened to mitigate security concerns.

In conclusion, the incorporation of a software TPM into the Firecracker VMM does not have a significant performance decrease. 
Instead, the extended use case, which features additional trust capabilities and assurance, compensates for the slight deterioration in performance.

\section*{Acknowledgments}
This work was partially supported by the Wallenberg AI, Autonomous Systems and Software Program (WASP) funded by the Knut and Alice Wallenberg Foundation. We would also like to extend our gratitude to Scionova AB through Erik Dahlgren and Daniel Hemberg, for their invaluable technical and financial support.

\printbibliography

@inproceedings{luo2019container,
  author    = {Wu Luo and Qingni Shen and Yutang Xia and Zhonghai Wu},
  title     = {{Container-IMA}: A privacy-preserving Integrity Measurement Architecture for Containers},
  booktitle = {22nd International Symposium on Research in Attacks, Intrusions and Defenses (RAID)},
  year      = {2019},
  isbn      = {978-1-939133-07-6},
  pages     = {487--500},
  publisher = {USENIX Association}
}

@inproceedings{firecracker-paper,
  author    = {Alexandru Agache and
               Marc Brooker and
               Alexandra Iordache and
               Anthony Liguori and
               Rolf Neugebauer and
               Phil Piwonka and
               Diana{-}Maria Popa},
  title     = {Firecracker: Lightweight Virtualization for Serverless Applications},
  booktitle = {17th {USENIX} Symposium on Networked Systems Design and Implementation (NSDI)},
  pages     = {419--434},
  year      = {2020}
}

@misc{cloud-hypervisor,
  title        = {{Implement vTPM support - Issue \#2343 - cloud-hypervisor/cloud-hypervisor}},
  howpublished = {\url{https://github.com/cloud-hypervisor/cloud-hypervisor/issues/2343}},
  note         = {Accessed: 2022-11-28}
}

@inproceedings{surbiryala2019cloud,
  title     = {Cloud computing: History and overview},
  author    = {Surbiryala, Jayachander and Rong, Chunming},
  booktitle = {2019 IEEE Cloud Summit},
  pages     = {1--7},
  doi       = {10.1109/CloudSummit47114.2019.00007}
}

@inproceedings{vtpm,
  author    = {Berger, Stefan and C\'{a}ceres, Ram\'{o}n and Goldman, Kenneth A. and Perez, Ronald and Sailer, Reiner and van Doorn, Leendert},
  title     = {{VTPM}: Virtualizing the Trusted Platform Module},
  year      = {2006},
  booktitle = {Proceedings of the 15th Conference on USENIX Security Symposium},
  articleno = {21}
}

@inproceedings{anjali_blending_2020,
  title      = {Blending containers and virtual machines: a study of {Firecracker} and {gVisor}},
  doi        = {10.1145/3381052.3381315},
  shorttitle = {Blending containers and virtual machines},
  eventtitle = {{VEE} '20: 16th {ACM} {SIGPLAN}/{SIGOPS} International Conference on Virtual Execution Environments},
  pages      = {101--113},
  booktitle  = {Proceedings of the 16th {ACM} {SIGPLAN}/{SIGOPS} International Conference on Virtual Execution Environments},
  author     = {{Anjali} and Caraza-Harter, Tyler and Swift, Michael M.},
  year       = {2020},
  langid     = {english}
}

@inproceedings{vtpm-vm-migration,
  author    = {Boris Danev and Ramya Jayaram Masti and Ghassan O. Karame and Srdjan Capkun},
  title     = {Enabling Secure {VM-VTPM} Migration in Private Clouds},
  year      = {2011},
  %isbn      = {9781450306720},
  publisher = {Association for Computing Machinery (ACM)},
  doi       = {10.1145/2076732.2076759},
  booktitle = {Proceedings of the 27th Annual Computer Security Applications Conference (ACSAC)},
  pages     = {187–196},
  numpages  = {10}
}

@inproceedings{sailer_ima_2004,
  author    = {Sailer, Reiner and Zhang, Xiaolan and Jaeger, Trent and van Doorn, Leendert},
  title     = {Design and Implementation of a {TCG}-Based Integrity Measurement Architecture},
  year      = {2004},
  booktitle = {Proceedings of the 13th Conference on USENIX Security Symposium - Volume 13},
  pages     = {223--238}
}

@article{AMIbrahim2019,
  doi       = {10.1016/j.cose.2018.12.014},
  year      = {2019},
  publisher = {Elsevier {BV}},
  pages     = {196--226},
  author    = {Fady A. M. Ibrahim and Elsayed E. Hemayed},
  title     = {Trusted Cloud Computing Architectures for infrastructure as a service: Survey and systematic literature review},
  journal   = {Computers \& Security}
}

@software{cloud_hypervisor_2023,
  title     = {{Cloud Hypervisor}},
  rights    = {Apache-2.0},
  url       = {https://github.com/cloud-hypervisor/cloud-hypervisor},
  publisher = {Cloud Hypervisor},
  urldate   = {2023-03-16},
  date      = {2019-04-30}
}

@software{crosvm_2023,
  title   = {crosvm},
  rights  = {{BSD}-3},
  url     = {https://chromium.googlesource.com/crosvm/crosvm/},
  urldate = {2023-03-16}
}

@inproceedings{bellard_qemu_2005,
  author    = {Bellard, Fabrice},
  title     = {{QEMU}, a Fast and Portable Dynamic Translator},
  year      = {2005},
  booktitle = {Proceedings of the Annual Conference on USENIX Annual Technical Conference},
  pages     = {41},
  numpages  = {1}
}

@misc{svtpm_a_2019,
  title      = {{SvTPM}: A Secure and Efficient {vTPM} in the Cloud},
  doi        = {10.48550/arXiv.1905.08493},
  shorttitle = {{SvTPM}},
  publisher  = {{arXiv}},
  author     = {Juan Wang and Chengyang Fan and Jie Wang and Yueqiang Cheng and Yinqian Zhang and Wenhui Zhang and Peng Liu and Hongxin Hu},
  year       = {2019}
}

@inproceedings{jia_hyperenclave_2022,
  title      = {{HyperEnclave}: An Open and Cross-platform Trusted Execution Environment},
  eventtitle = {2022 {USENIX} Annual Technical Conference (ATC)},
  pages      = {437--454},
  author     = {Jia, Yuekai and Liu, Shuang and Wang, Wenhao and Chen, Yu and Zhai, Zhengde and Yan, Shoumeng and He, Zhengyu},
  langid     = {english}
}

@inproceedings{pecholt_cocotpm_2022,
  title     = {{CoCoTPM}: Trusted Platform Modules for Virtual Machines in Confidential Computing Environments},
  doi       = {10.1145/3564625.3564648},
  pages     = {989--998},
  booktitle = {Proceedings of the 38th Annual Computer Security Applications Conference (ACSAC)},
  author    = {Joana Pecholt and Sascha Wessel},
  year      = {2022}
}

@online{mitre_corporation_cve_2023,
  title   = {{CVE} - {CVE}-2023-1018},
  url     = {https://cve.mitre.org/cgi-bin/cvename.cgi?name=CVE-2023-1018},
  author  = {{MITRE Corporation}},
  urldate = {2023-03-27},
  date    = {2023-02-24}
}

@online{falcon_vulnerabilities_2023,
  title   = {Vulnerabilities in the {TPM} 2.0 reference implementation code},
  url     = {https://blog.quarkslab.com/vulnerabilities-in-the-tpm-20-reference-implementation-code.html},
  author  = {Falcon, Francisco},
  urldate = {2023-03-27},
  date    = {2023-03-14}
}

@inproceedings{manco2017my,
  title     = {My {VM} is Lighter (and Safer) than your Container},
  author    = {Manco, Filipe and Lupu, Costin and Schmidt, Florian and Mendes, Jose and Kuenzer, Simon and Sati, Sumit and Yasukata, Kenichi and Raiciu, Costin and Huici, Felipe},
  booktitle = {Proceedings of the 26th Symposium on Operating Systems Principles},
  doi       = {10.1145/3132747.3132763},
  pages     = {218--233},
  year      = {2017}
}

@inproceedings{young2019true,
  title     = {The True Cost of Containing: A gVisor Case Study.},
  author    = {Young, Ethan G and Zhu, Pengfei and Caraza-Harter, Tyler and Arpaci-Dusseau, Andrea C and Arpaci-Dusseau, Remzi H},
  booktitle = {Workshop on Hot Topics in Cloud Computing (HotCloud)},
  year      = {2019}
}

@inproceedings{morabito_hypervisors_2015,
  title      = {Hypervisors vs. Lightweight Virtualization: A Performance Comparison},
  doi        = {10.1109/IC2E.2015.74},
  shorttitle = {Hypervisors vs. Lightweight Virtualization},
  eventtitle = {2015 {IEEE} International Conference on Cloud Engineering},
  pages      = {386--393},
  booktitle  = {2015 {IEEE} International Conference on Cloud Engineering},
  author     = {Morabito, Roberto and Kjällman, Jimmy and Komu, Miika}
}

@inproceedings{tesfatsion_virtualization_2018,
  title      = {Virtualization Techniques Compared: Performance, Resource, and Power Usage Overheads in Clouds},
  doi        = {10.1145/3184407.3184414},
  shorttitle = {Virtualization Techniques Compared},
  pages      = {145--156},
  booktitle  = {Proceedings of the 2018 {ACM}/{SPEC} International Conference on Performance Engineering (ICPE)},
  author     = {Tesfatsion, Selome Kostentinos and Klein, Cristian and Tordsson, Johan},
  year       = {2018}
}

@article{chae_performance_2019,
  title        = {A performance comparison of {Linux} containers and virtual machines using {Docker} and {KVM}},
  doi          = {10.1007/s10586-017-1511-2},
  pages        = {1765--1775},
  journaltitle = {Cluster Computing},
  author       = {Chae, {MinSu} and Lee, {HwaMin} and Lee, Kiyeol},
  year         = {2019},
  langid       = {english}
}

@inproceedings{felter_updated_2015,
  title     = {An updated performance comparison of virtual machines and {Linux} containers},
  doi       = {10.1109/ISPASS.2015.7095802},
  pages     = {171--172},
  booktitle = {2015 {IEEE} International Symposium on Performance Analysis of Systems and Software ({ISPASS})},
  author    = {Felter, Wes and Ferreira, Alexandre and Rajamony, Ram and Rubio, Juan},
  date      = {2015-03}
}

@article{russell_virtio_2008,
  title        = {virtio: towards a de-facto standard for virtual I/O devices},
  %issn         = {0163-5980},
  doi          = {10.1145/1400097.1400108},
  pages        = {95--103},
  journaltitle = {{ACM} {SIGOPS} Operating Systems Review},
  shortjournal = {{SIGOPS} Oper. Syst. Rev.},
  author       = {Russell, Rusty},
  date         = {2008-07-01},
  keywords     = {I/O, {KVM}, lguest, Linux, ring buffer, virtio, virtio\_pci, virtualization, vring}
}

@misc{cahalan_pmap1_nodate,
  title        = {pmap(1): report memory map of process - Linux man page},
  howpublished = {\url{https://linux.die.net/man/1/pmap}},
  note         = {Accessed: 2023-04-25}
}

@inproceedings{duplyakin_design_cloudlab_2019,
  title     = {The Design and Operation of {CloudLab}},
  pages     = {1--14},
  booktitle = {Proceedings of the {USENIX} Annual Technical Conference (ATC)},
  author    = {Duplyakin, Dmitry and Ricci, Robert and Maricq, Aleksander and Wong, Gary and Duerig, Jonathon and Eide, Eric and Stoller, Leigh and Hibler, Mike and Johnson, David and Webb, Kirk and Akella, Aditya and Wang, Kuangching and Ricart, Glenn and Landweber, Larry and Elliott, Chip and Zink, Michael and Cecchet, Emmanuel and Kar, Snigdhaswin and Mishra, Prabodh},
  year      = {2019}
}

@inproceedings{szefer_eliminating_2011,
  title     = {Eliminating the hypervisor attack surface for a more secure cloud},
  %isbn      = {978-1-4503-0948-6},
  doi       = {10.1145/2046707.2046754},
  pages     = {401--412},
  booktitle = {Proceedings of the 18th {ACM} conference on Computer and communications security},
  author    = {Szefer, Jakub and Keller, Eric and Lee, Ruby B. and Rexford, Jennifer},
  year      = {2023},
  langid    = {english}
}

@misc{tolnay_chromium_nodate,
  title        = {{CHROMIUM}: Device driver for {TPM} over virtio (1480209) - {Gerrit} Code Review},
  howpublished = {\url{https://chromium-review.googlesource.com/c/chromiumos/third_party/kernel/+/1480209/2}},
  note         = {Accessed: 2023-04-26}
}

@software{aldanial_cloc_2022,
  title  = {{AlDanial}/cloc: v1.96},
  year   = {2022},
  doi    = {10.5281/zenodo.7455676},
  author = {{AlDanial} and Snel, Sietse and Boos, Stefan and {jolkdarr} and Beckmann, Christoph and Dimmitt, Michael and Wilk, Jakub and Chaves, Gustavo and Rudis BoB and {asrmchq} and Gough, Alex and Tang, Jimmy and Dursi, Jonathan and {RyanMcC} and {achary} and Ali, Asad and {Brando!} and Dahlheimer, Chris and Losantos, David and  Ulrich, David and {erkmos} and Brinkhoff, Lars and {LoganDark} and Irländer, Torsten and Rösler, Wolfram and  Long, Fei and {b1f6c1c4} and Solanki, Vibhakar and HOUZÉ, Sébastien and Ryan, Alex}
}

@misc{zhang_top_2023,
  title    = {Top 10 {Cloud} {Service} {Providers} {Globally} in 2023},
  url      = {https://dgtlinfra.com/top-10-cloud-service-providers-2022/},
  language = {en-US},
  urldate  = {2023-06-12},
  journal  = {Dgtl Infra},
  author   = {Zhang, Mary},
  month    = jan,
  year     = {2023}
}

\end{document}